\documentclass[
aps,%
12pt,%
final,%
notitlepage,%
oneside,%
onecolumn,%
nobibnotes,%
nofootinbib,%
superscriptaddress,%
noshowpacs,%
centertags]%
{revtex4}

\begin{document}

\title{Detection of Gravitational Waves through Observations\\
of a Group of Pulsars}

\author{\firstname{A.~E.}~\surname{Rodin}}

\email{rodin@prao.ru}
\affiliation{%
Pushchino Radio Astronomy Observatory, Astro Space Center, Lebedev Physical Institute, Russian Academy of Sciences
}%


\begin{abstract}
We suggest a new approach to the detection of gravitational waves using observations of a group of millisecond pulsars. In contrast to the usual method, based on increasing the accuracy of the arrival times of pulses by excluding possible distorting factors, our method supposes that the additive phase noise that is inevitably present even in the most accurate observational data has various spectral components, which have characteristic amplitudes and begin to appear on different time scales. We use the ``Caterpillar'' (Singular Spectral Analysis, SSA) method to decompose the signal into its components. Our initial data are the residuals of the pulse arrival times for six millisecond pulsars. We constructed the angular correlation function for components of the decomposition of a given number, whose theoretical form for the case of an isotropic and homogeneous gravitational-wave background is known. The individual decomposition components show a statistically significant agreement with the theoretical expectations  (correlation coefficient $\rho=0.92\pm0.10$).
\end{abstract}

\maketitle

{\bf DOI}: 10.1134/S1063772911020041

\section{Introduction}

Long-term analyses of pulsar-timing data have shown that the fractional instability of the rotations of some pulsars are comparable to the instabilities of atomic frequency standards, and can reach ${\Delta\nu}/{\nu}\sim 10^{-15}$. This makes it possible to apply timing data to various astronomical and metrological problems. In the present paper, we describe a method for detecting gravitational waves using observations of a group of millisecond pulsars. The method is based on the so-called two-point (angular) correlation function \cite{HellingsDowns1983, ZhaoZhang2003,jenet2005}
\begin{equation}\label{two-point}
\zeta(\theta)=\frac32x\log(x)-\frac x4+\frac12+\frac12\delta(x),
\end{equation}
where $x=(1-\cos\theta)/2$, $\theta$ is the angular separation of the pulsars. After the publication of \cite{jenet2005} the principle of gravitational-wave detection using pulsar-timing data became clear, and observations employing this method were initiated in many observatories.

The main attention in these programs was focused on increasing the accuracy of pulse time-of-arrival (TOA) determinations and excluding all possible distorting physical factors (instability of the reference clock, influence of the interstellar medium, the intrinsic activity of the pulsars themselves, etc.). However, it is clear that increasing the accuracy of the observations is a necessary, but not sufficient, condition, since increasingly finer effects are found, which have a stochastic nature and are able to influence the gravitational-wave background.

We suggest here a fundamentally different approach to detecting gravitational waves. It is known \cite{petit1996} that physical objects such as atomic time standards or pulsars display noise over a broad frequency range. For example, the phase noise of a frequency standard or a pulsar contains both white noise, which is usually identified with errors of the detectors, and correlated (red) noise with various spectral indexes \cite{blandford1984}, which begins to appear in data collected on various time intervals. Moreover, the analysis of time series corresponding to the different types of noises shows that they have characteristic features that can be used to discriminate between them. These features suggest that there should exist a method for decomposing time series into components having different spectral compositions. If one such component is dominated by the gravitational-wave background, the presence of this background should be distinguishable in the angular correlation function. We choose the ``Caterpillar'' or Singular Spectral Analysis (SSA) method \cite{golyandina2004} to decompose the time series into separate components. This method is functionally independent, since the time series itself is used to obtain the orthogonal basis functions.

\section{Observations}

Observations of the pulsars PSR J0613-0200, J1640+2224, J1643-1224, J1713+0747, J1939+2134, and J2145-0750 were carried out using the fully steerable 64-m radio telescope of the Kalyazin Radio Astronomical Observatory of the Astro Space Center of the Lebedev Physical Institute \cite{potapov2003,ilyasov2004a,ilyasov2004b,ilyasov2005}. The pulses were accumulated in two circular polarizations using a spectrum analyzer with an 80-channel filter bank with a bandwidth of 40kHz per channel in each polarization \cite{oreshko2000}. Each pulsar was observed, on average, once in two weeks. The total signal sampling time was close to two hours per session. The pulse TOAs were determined using a local time standard with accuracy better than 100 ns. The local time standard was tied to UTC(SU) via a television channel and to UTC(USNO) via a GPS receiver. The pulses were accumulated in three-minute cycles synchronous with the pulsar rotation, with the subsequent restart of the system using the newly computed observed pulsar period. The data for each three-minute cycle were recorded in a separate file. The pulse delays due to frequency dispersion in the individual channels were compensated for in the subsequent data reduction.

The topocentric pulse TOAs were determined by matching the pulsar profile summed over a session to a standard high signal-to-noise profile (template). The barycentric pulse TOAs, TOA residuals, and refined astrometric and rotational parameters of the pulsars were calculated via a least-squares minimization of the TOA residuals using the Tempo package \cite{Taylor1989}. The DD model \cite{Damour1986} was used to refine the orbital parameters of the pulsars. We used the values given in the pulsar catalog  \cite{Manchester2005} as the initial values.
 
Table 1 lists the least-squares-fit timing parameters for the pulsars: the right ascension $\alpha(J2000)$, declination $\delta(J2000)$, proper motions $\mu_\alpha$, $\mu_\delta$ (in mas/yr) for epoch J2000, the rotational frequency $f$ and its derivative  $\dot f$ (in s$^{-1}$ and s$^{-2}$ ), the dispersion measure $DM$  (in pc/cm$^3$ ), the projection of the semi-major axis of the pulsar orbit onto the line of sight $x$ (in light seconds), the eccentricity of the orbit $e$, the epoch of the pulsar's transit  $T_\Pi$ (in MJD), the longitude of the orbit periastron $\omega$ (in deg), the rms error of the residuals after improving the pulsar timing parameters $\sigma$ (in $\mu$s). Since the observations were carried out in a single-frequency regime, we did not correct for the $DM$ . The formal least-squares error of the last significant digit of the parameter is given in parentheses.

To compute the components of the pulse-TOA decompositions using the SSA method, the time-series were averaged over 40-day intervals, filling gaps via a linear interpolation between adjacent measurements. We took only common parts of the pulse-TOA series in the range MJD = 51000--53000. After averaging and interpolation, the pulse-TOA series contained $N = 51$ points. The series of pulse-TOA residuals averaged as described above are shown in Fig. 1. 

\section{ The ``Caterpillar'' (SSA) method}

We will now describe the steps in the SSA method for a one-dimensional time-series  $\{x_i\}_{i=1}^N$ with length $N$ following \cite{zhiglavski1997}.

\begin{enumerate}
\item The first step is unfolding the one-dimensional
series into a multi-dimensional series. We take a
number $M<N$ (the length of ``caterpillar''), specify $k=N-M+1$ and form the matrix 
$X=({x_{ij}})_{i,j=1}^{k,M}$ with elements $x_{ij}=x_{i+j-1}$.
\begin{equation}
X=({x_{ij}})_{i,j=1}^{k,M}=\left(
\begin{array}{ccccc}
x_1 & x_2 & x_3 & \ldots & x_M\\
x_2 & x_3 & x_4 & \ldots & x_{M+1}\\
x_3 & x_4 & x_5 & \ldots & x_{M+2}\\
\vdots & \vdots & \vdots & \ddots & \vdots \\
x_{k} & x_{k+1} & x_{k+2} &\ldots & x_N
\end{array}
\right).
\end{equation}

\item Next, the following matrix is solved:
\begin{equation}
R=XX^T,
\end{equation}
where $()^T$ denote conjugation.

\item The eigenvalues and eigenvectors of the matrix $R$ are then computed, i.e., we find expand it in the form
\begin{equation}
R=P\Lambda P^T,
\end{equation}
where
\begin{equation}
\Lambda=\left(
\begin{array}{cccc}
\lambda_1 & 0 & \dots & 0\\
0 & \lambda_2 & \dots & 0\\
\vdots & \vdots & \ddots & \vdots \\
0 & 0 & \dots & \lambda_M\\
\end{array}
\right)
\end{equation}
is a diagonal matrix of the eigenvalues and
\begin{equation}
P=(p_1,p_2,p_3,\ldots,p_M)=\left(
\begin{array}{cccc}
p_{11} & p_{21} & \dots & p_{M1}\\
p_{12} & p_{22} & \dots & p_{M2}\\
\vdots & \vdots & \ddots & \vdots \\
p_{1M} & p_{2M} & \dots & p_{MM}\\
\end{array}
\right)
\end{equation}
is an orthogonal matrix of the eigenvectors of the
matrix $R$. Note that \cite{zhiglavski1997}: $P^T=P^{-1}$, $P^TP=PP^T=I_M$,
$\Lambda=P^TRP$, $\sum_{i=1}^M\lambda_i=M$, $\prod_{i=1}^M\lambda_i={\rm det}R$,
where $I_M$ is the identity matrix with dimension $M$ and ${\rm det(\cdot)}$ is the determinant of the matrix  $(\cdot)$.

\item The transformation to the principal components is accomplished using the formula
\begin{equation}
XP=Y=(y_1,y_2,\ldots,y_M),
\end{equation}
and the inverse transformation to the matrix $X$ is computed $X=YP^T$.

\item Further is the reconstruction of the initial series.
The reconstruction procedure is based on the formula $\tilde X=Y^*P^T$. 
It is said that the reconstruction is carried out using a given set of principal components if the matrix Y* is obtained from the matrix Y by setting all components not included in the set of principle components to zero. Thus, we can obtain an approximation for the matrix of interest, or to some interpretable part of this matrix. The one-dimensional time-series is obtained via a diagonal averaging of  $\tilde X$ using the formulas:
\begin{equation}
\tilde x_s=
\left\{
\begin{array}{lr}
\frac 1s\sum\limits_{i=1}^s\tilde x_{i,s-i+1}, & 1\leq s \leq M,\\
\frac 1M\sum\limits_{i=1}^M\tilde x_{i,s-i+1}, & M\leq s \leq k,\\
\frac 1{N-s+1}\sum\limits_{i=1}^{N-s+1}\tilde x_{i+s-k,k-i+1}, & k\leq s \leq N.
\end{array}
\right.
\end{equation}

\end{enumerate}

\section{Mathematical modelling}

We computed a mathematical model to analyse the accuracy of the component reconstruction using the SSA method. We took the pulse-TOA residuals of the pulsars PSR J0613-0200, J1640+2224, J16431224, J1713+0747, J1939+2134, and J2145-0750 as the initial reconstructed series. We added to these a series of normally distributed random values with zero mean and a dispersion based on the instrumental error in the pulse TOAs. In total, we generated 200 random realizations for each pulsar. The instrumental errors were taken from the paper of Oreshko \cite{oreshko2000}, who presents the errors of the pulse TOAs determined using the AS-600 installation. The main input to the error is made by instability of the frequency of the bandpass filter. For the AS-600, this is given by the formula  $\sigma_t(f)=0.0073\,DM\; \mu$s. Additional sources of inaccuracy are the local time standard ($\sim 100$ ns) and the period synthesizer (($\sim 10$ ns) and recorder ($\sim 20$ ns). Table 2 gives the instrumental errors  $\sigma_\tau$ for the pulse TOAs for pulsars with various $DM$s and the mean error of each decomposition component $\sigma_1$ .

Next, the series of pulse TOAs were decomposed into components using the SSA method and the rms difference between the reconstructed and initial components was calculated. Figure 2 shows the results of our modelling for each of the six pulsars. The rms deviations are not uniformly distributed between the components -- on average, components with higher numbers have lower errors. The error in the reconstructed components comprises 19\%-32\% of the instrumental error. 

We computed another mathematical model to analyse the real accuracy of the angular correlation function (\ref{two-point}). Six pulsars were randomly placed on
the celestial sphere in a Cartesian coordinate system Oxyz \cite{HellingsDowns1983}. Next, the relative variation of the pulse frequency was calculated according to the relative positions of the pulsar and the gravitational wave, and the angular correlation coefficient $\zeta(\theta)$ was calculated. In all, $N = 51$ trials were made, in accordance with the number of data for each pulsar. Figure 3 shows the two-point correlation function obtained from these calculations averaged over 0.13 rad. The random error in the correlation coefficient is $\sigma_\rho=0.15$.

\section{Results}

To aid a visual analysis, we artificially added a seventh ``pulsar'' with the mean residuals \cite{rodin2008} and the mean coordinates of the six pulsars. This does not add new data, but improves the filling of the plot.

Figure 4 shows the component decomposition of the pulse-TOA residuals using the SSA method. The decrease in the period and amplitude of the components as the component number increases is clearly visible. We calculated power spectra (periodograms) for each component, shown in Fig. 5. An interesting feature of these spectra is that components with low numbers (1-5) display a well defined rise at low frequencies, corresponding to the presence of so-called ``red'' noises. Components 6-9 display a maximum in the middle of the frequency range, while components 10-12 rise at high frequencies (so-called ``blue'' spectra). Thus, in terms of spectral indices, the spectral index of the periodogram grows as the component number increases. Since, generally speaking, spectra with variable spectral indices can be associated with noise having various physical natures \cite{blandford1984}, we conclude that the SSA method for decomposing the time series into components enables us to distinguish components having various physical origins.

We calculated the angular correlation functions given by (1) for the decomposition components of a given number, shown in Fig. 6. The experimental two-point correlation functions for components 8 and 9 shows a good agreement with the theoretical function: the correlation coefficient between the theoretical and experimental points is $\rho=0.92\pm0.10$. For this correlation coefficient and the number of points N = 19, the probability that this correlation occurred by chance is ${\rm Pr}(\rho=0.92,N=19)\lesssim 10^{-7}$ \cite{jw1971}. The
two-point correlation function for the 8th component smoothed over two to five points is shown in more detail in Fig. 7.

There are several possible explanations for this result. The most interesting would be the actual detection of gravitational waves. If the correlated amplitude of the 8th and 9th components of the decomposition, equal to 0.5 $\mu$s, is recalculated to the energy density of the inferred gravitational-wave background taking into account the observation span of six years \cite{kaspi1994}, we obtain $\Omega_g h^2\sim 10^{-8}$, which is much higher than previous upper limits for the gravitational-wave background obtained in many studies (see, e.g., \cite{lorimer2008}), which give  $\Omega_g h^2 < 10^{-10}$.

It is also possible that this result is associated with some as yet unknown process, which behaves similarly to a gravitational-wave background with respect to the angular correlation function, and which requires additional study.

\section{Conclusion}

We have suggested a new method for detecting a gravitational-wave background based on observations of a group of millisecond pulsars. The method employs a new approach, based on a preliminary decomposition of the pulse TOA residuals of pulsars into spectral components with various physical natures and their subsequent analysis using the angular correlation function. We have applied this approach to the pulse-TOA residuals for the six pulsars PSR J0613-0200, J1640+2224, J1643-1224, J1713+0747, J1939+2134, and J2145-0750. This has yielded a calculated angular cross-correlation function that agrees with the function derived for the gravitational-wave background at a statistically significant level.

\begin{acknowledgments}
The author thanks the staff of the Department of Plasma Astrophysics of the Pushchino Radio Astronomy Observatory for fruitful discussions and suggestions. This work was partially supported by the Russian Foundation for Basic Research (project 09-02-00584-a).
\end{acknowledgments}

\newpage

\pagebreak
\begin{table}\label{tab1}
\setcaptionmargin{0mm} \onelinecaptionsfalse
\captionstyle{flushleft} 
\caption{Estimates of the timing parameters for the six pulsars}
\bigskip
{\footnotesize
\begin{tabular}{|l|c|c|c|c|c|c|}
\hline
Parameter & J0613--0200 & J1640+2224 & J1643--1224 & J1713+0747 & J1939+2134 &
J2145--0750 \\
\hline
$\alpha(J2000)$&${06:13:43\atop .974036(98)}$&${16:40:16\atop .743755(36)}$
&${16:43:38\atop .155649(83)}$&${17:13:49\atop .527445(41)}$
&${19:39:38\atop .5612474(65)}$&${21:45:50\atop .4677(22)}$\\
$\delta(J2000)$&${-02:00:47\atop
.0662(43)}$&${22:24:09\atop .01404(63)}$&${-12:24:58\atop .7356(49)}$
&${07:47:37\atop .5373(11)}$&${21:34:59\atop
.13479(12)}$&${-07:50:18\atop .2665(90)}$\\
$\mu_\alpha(J2000)$&1.9(2)&0.02(6)&6.0(2)&4.91(6)&-0.003(15)&-1.5(5)\\
$\mu_\delta(J2000)$&-10.4(6)&-10.60(9)&2.7(9)&-1.6(1)&-0.75(2)&-26.6(1.5)\\
$f,\;s^{-1}$&${326\atop
.600567572653(72)}$&${316\atop .1239843704318(34)}$&${216\atop
.373340906436(41)}$&${218\atop .8118440285461(29)}$&${641.\atop
928252901190(15)}$&${62\atop .2958888585892(97)}$\\
$\dot f,\;10^{-16}s^{-2}$
&-10.232(3)&-2.8119(1)&-8.6413(3)&-4.07952(9)&-431.1030(2)&-1.1538(2)\\
$DM$, pc/cm$^{3}$&38.785&18.415&62.404&15.993&71.040&9.000\\
$x$, light sec&1.091438(1)&55.329732(1)&25.072609(2)&32.342425(1)&& 10.16409(2)\\
$e$&0.0000000(6)&0.00079715(4)&0.0005053(2)&0.00007488(9)&& 0.0000183(3)\\
$T_\Pi$, MJD& ${49513\atop.1983545(2)}$&${49345\atop
.1573876(6)}$&${49577\atop .987029(2)}$&${48741\atop .9551306(5)}$&
&${49910\atop .1985342(2)}$\\
$P_b$, day&1.19807812205(9)&175.46256914(3)&147.0152985(1)&67.825130005(8)&
&6.8388737600(7)\\
$\omega$, degree&280.06527(6)&50.664174(1)&321.894409(5)&176.101042(2)&
&199.855700(9)\\
$\sigma$, $\mu$s &18.52&21.04&43.63&17.48&5.24&42.37\\
\hline
\end{tabular}}
\end{table}

\pagebreak
\begin{table}\label{tab2}
\setcaptionmargin{0mm} \onelinecaptionsfalse
\captionstyle{flushleft} 
\caption{ Instrumental errors in the pulse TOAs $\sigma_\tau$ as
functions of the $DM$ and the error in the reconstruction of
the decomposition components $\sigma_1$.}
\bigskip
\begin{tabular}{|l|c|c|c|}
\hline
Pulsar & $DM$ , pc/cm$^3$ & $\sigma_\tau$, $\mu$s & $\sigma_1$, $\mu$s \\
\hline
J0613--0200 & 38.785 & 0.30 & 0.064\\
J1640+2224  & 18.415 & 0.17 & 0.055\\
J1643--1224 & 62.404 & 0.47 & 0.095\\
J1713+0747  & 15.993 & 0.16 & 0.032\\
J1939+2134  & 71.040 & 0.53 & 0.14\\
J2145--0750 & 9.000  & 0.12 & 0.023\\
\hline
\end{tabular}
\end{table}

\begin{figure}\label{fig1}
\setcaptionmargin{5mm}
\onelinecaptionstrue
\includegraphics[width=1\textwidth]{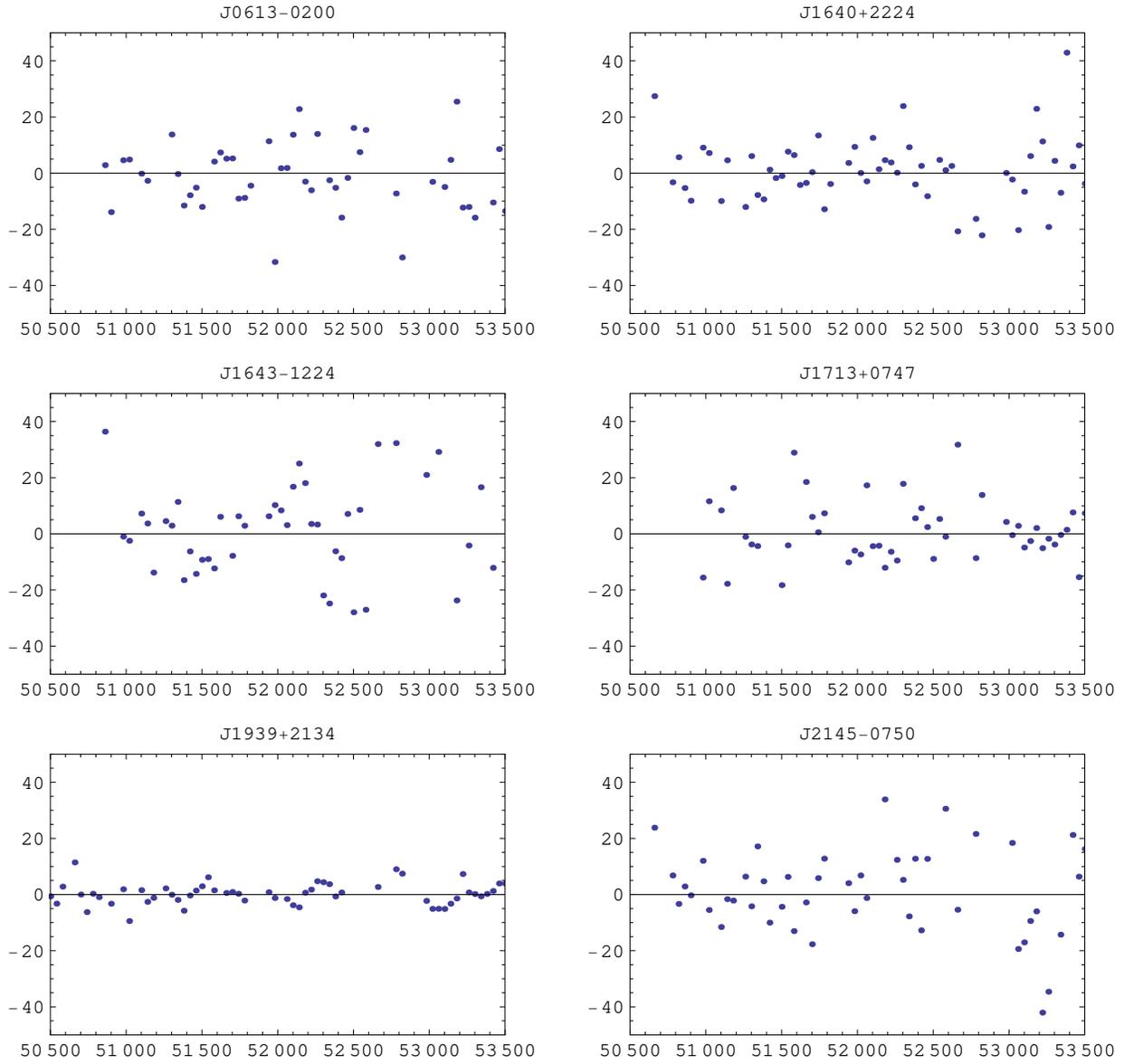}
\captionstyle{normal}
\caption{ Pulse-TOA residuals for the six pulsars averaged over 40 days (in $\mu$s). The modified Julian date is plotted along the horizontal axis.}
\end{figure}

\begin{figure}\label{fig2}
\setcaptionmargin{5mm}
\onelinecaptionstrue
\includegraphics[width=1\textwidth]{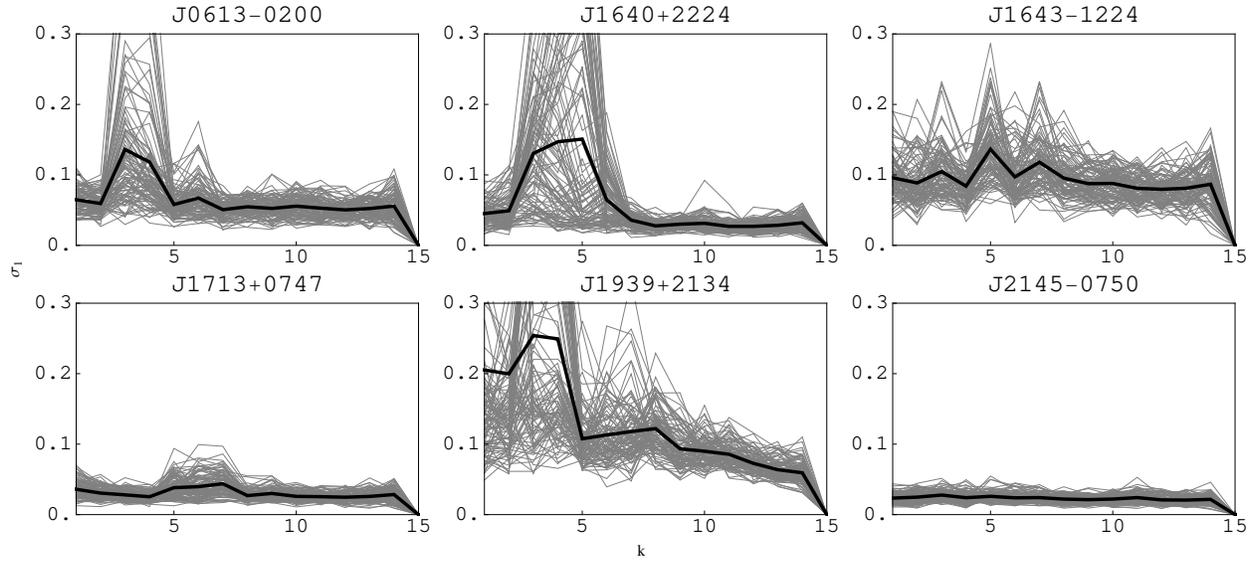} 
\captionstyle{normal}
\caption {Error in the reconstruction $\sigma_1$ (y axis) in $\mu$s as a function of the number of decomposition components $k$ (x axis) for the six pulsars.}
\end{figure}

\begin{figure}\label{cor-sim}
\setcaptionmargin{5mm}
\onelinecaptionstrue
\includegraphics[width=1\textwidth]{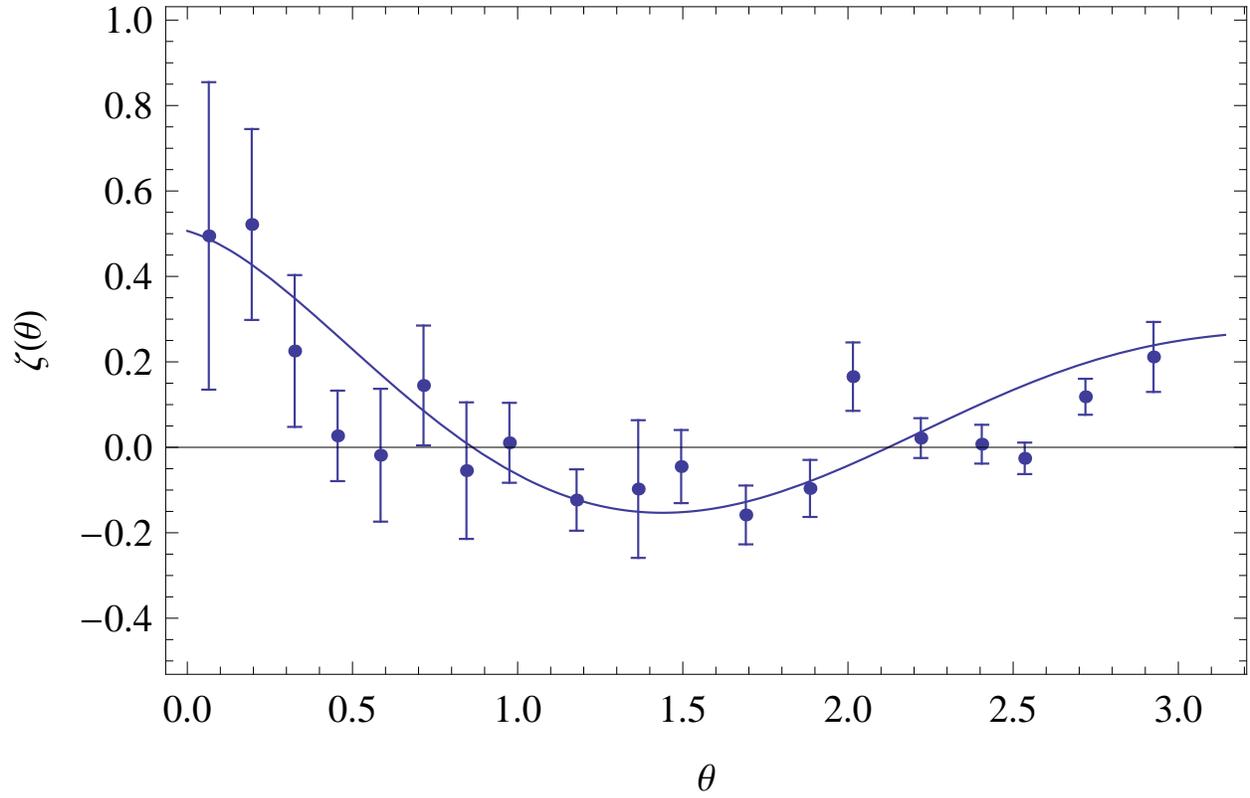}
\captionstyle{normal}
\caption {Results of numerical modelling for the two-point correlation function. The data for the six pulsars with $N = 51$ pulse TOAs is used.}
\end{figure}

\pagebreak

\begin{figure}\label{psr-gus}
\setcaptionmargin{5mm}
\onelinecaptionstrue
\includegraphics[width=0.95\textwidth]{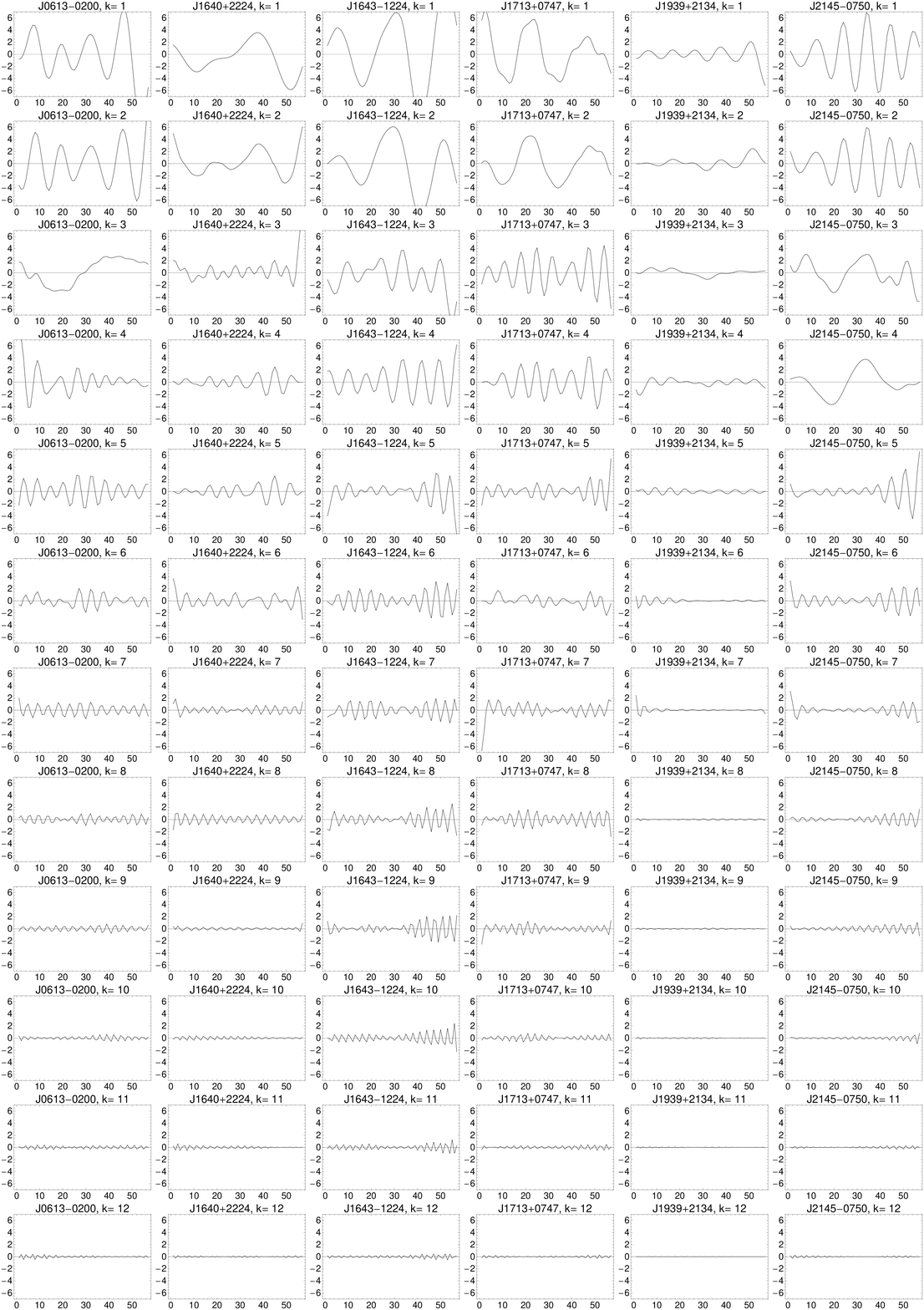}
\captionstyle{normal}
\caption{Decomposition of TOA residuals into components for the six pulsars using the SSA method. The pulsar number and component number are indicated above each plot. The horizontal axis plots the time in 40-day intervals, and the vertical axis the time in $\mu$s.}
\end{figure}

\pagebreak

\begin{figure}\label{psr-sp}
\setcaptionmargin{5mm}
\onelinecaptionstrue
\includegraphics[width=0.95\textwidth]{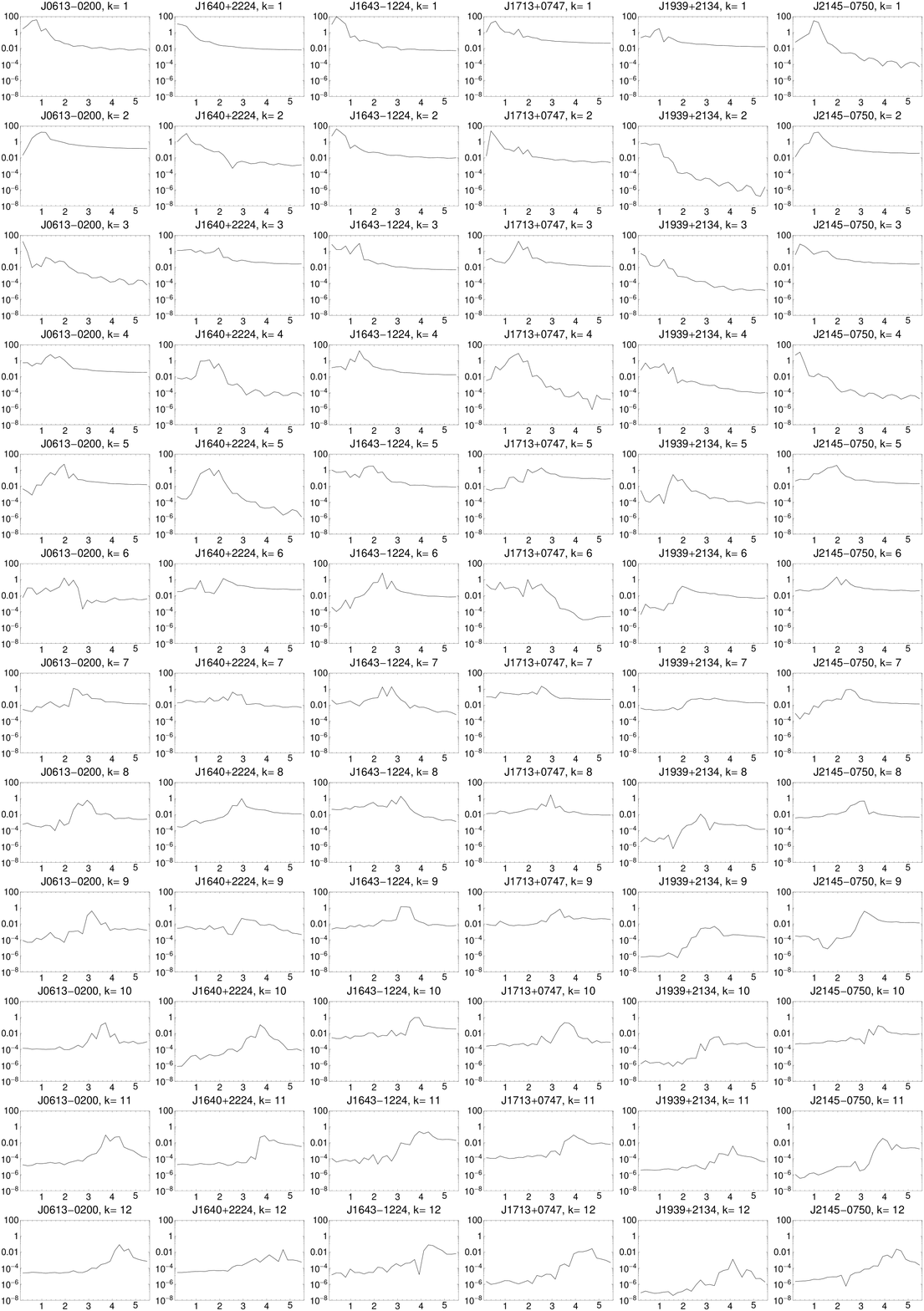}
\captionstyle{normal}
\caption{ Power spectra of the TOA-residual decomposition components for the six pulsars obtained using the SSA method. The pulsar number and component number are indicated above each plot. The horizontal axis plots the frequency (in yr$^{-1}$ ) and the vertical axis the power (in $\mu{\rm s}^2$).}
\end{figure}

\begin{figure}\label{cor-psr}
\setcaptionmargin{5mm}
\onelinecaptionstrue
\includegraphics[width=1\textwidth]{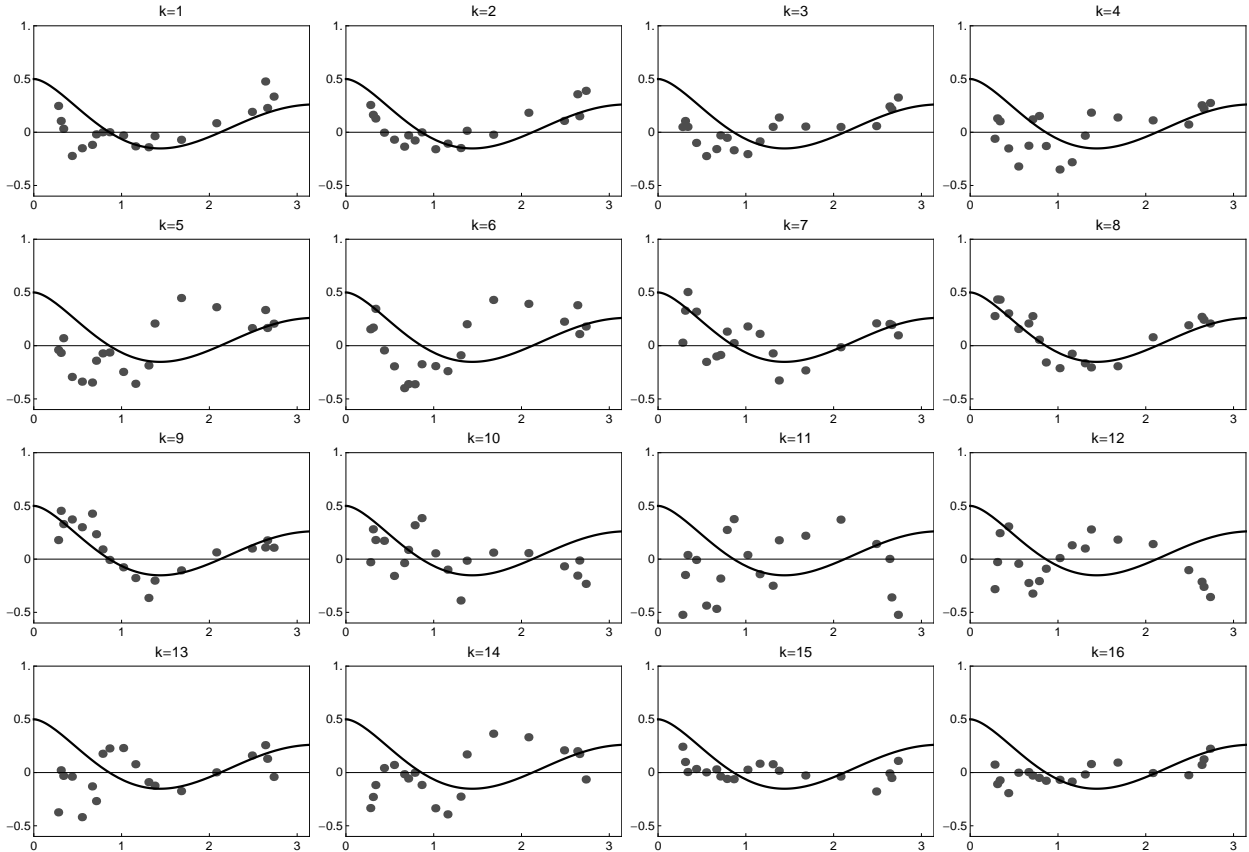}
\captionstyle{normal}
\caption{ Angular-correlation function calculated for the various TOA-residual decomposition components for the six pulsars smoothed over three points (the component numbers are indicated above the plots). The solid curves show the theoretical dependencies calculated assuming the presence of a gravitational-wave background. Components 8 and 9 show statistically significant correlations with the theoretical curves.}
\end{figure}

\begin{figure}\label{cor-av}
\setcaptionmargin{5mm}
\onelinecaptionstrue
\includegraphics[width=1\textwidth]{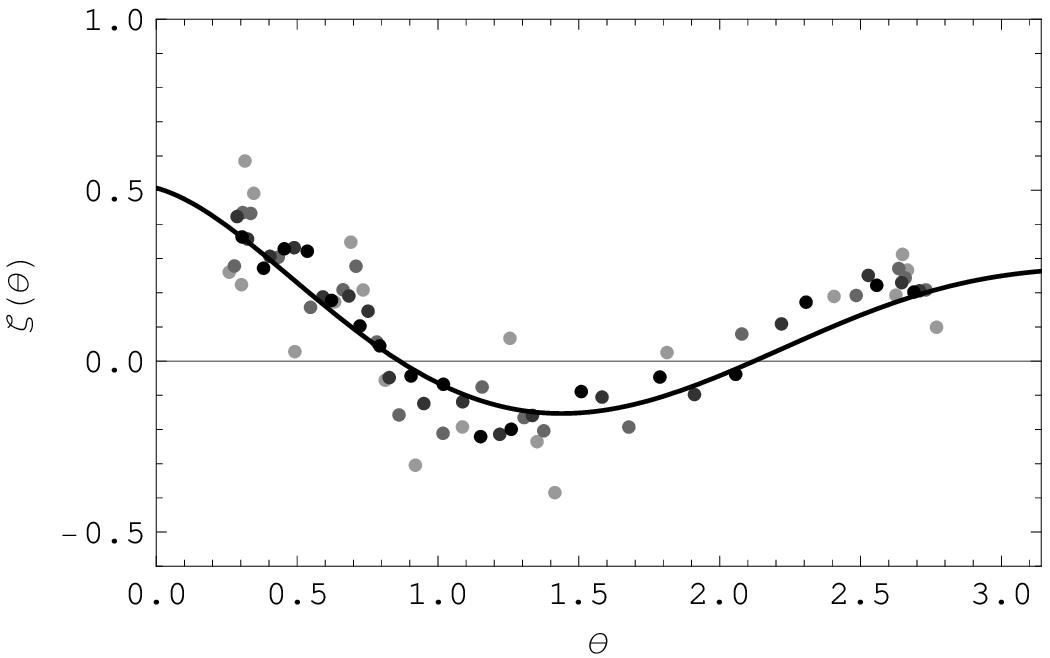}
\captionstyle{normal}
\caption{ Angular-correlation function calculated for the eighth component of the TOA-residual decomposition for the six pulsars smoothed over two to five points. The solid curve shows the theoretical dependence calculated assuming the presence of a gravitational-wave background.}
\end{figure}


\begin{thebibliography}{99}
\bibitem{jenet2005} F. A. Jenet, G. B. Hobbs, K. J. Lee, and R. N. Manchester, Astrophys. J. 625, L123 (2005).
\bibitem{HellingsDowns1983} R. W. Hellings and G. S. Downs, Astrophys. J. 265, L39 (1983).
\bibitem{ZhaoZhang2003} W. Zhao and Y. Zhang, Acta Astron. Sinica 44S, 273 (2003).
\bibitem{petit1996} G. Petit and P. Tavella, Astron. Astrophys. 309, 290 (1996).
\bibitem{blandford1984} R. Blandford, R. Narayan, and R. Romani, J. Astrophys. Astron. 5, 369 (1984).
\bibitem{golyandina2004} N. E. Golyandina, The Caterpillar-SSA Method: Time Series Analysis . A Textbook (SPb Gos. Univ., St. Petersburg, 2004) [in Russian].
\bibitem{ilyasov2004a} Y. P. Ilyasov, V. V. Oreshko, V. A. Potapov, and A. E. Rodin, ASP Conf. Ser. 218, 433 (2004).
\bibitem{ilyasov2004b} Y. P. Ilyasov, M. Imae, Y. Hanado, et al., ASP Conf. Ser. 218, 435 (2004).
\bibitem{ilyasov2005} Yu. P. Ilyasov, M. Imae, Yu. Hanado, et al., Pis'ma Astron. Zh. 31, 33 (2005) [Astron. Lett. 31, 30 (2005)].
\bibitem{potapov2003} V. A. Potapov, Yu. P. Ilyasov, V. V. Oreshko, and A. E. Rodin, Pis'ma Astron. Zh. 29, 282 (2003) [Astron. Lett. 29, 241 (2003)].
\bibitem{oreshko2000} V. V. Oreshko, Tr. Fiz. Inst. Lebedeva 229, 110 (2000).
\bibitem{Taylor1989} J. H. Taylor and J. M. Weisberg, Astrophys. J. 345, 434 (1989).
\bibitem{Damour1986} D. Damour and N. Deruelle, Ann. Inst. Henri Poincare. Phys. Theor. 44, 263 (1986).
\bibitem{Manchester2005} R. N. Manchester, G. B. Hobbs, A. Teoh, and M. Hobbs, Astron. J. 129, 1993 (2005).
\bibitem{zhiglavski1997} The Main Components of Time Series: The ``Caterpillar'' Method, Ed. by D. L. Danilov and A. A. Zhiglyavskii (SPb. Gos. Univ., St. Petersburg, 1997) [in Russian].
\bibitem{rodin2008} A. E. Rodin, Mon. Not. R. Astron. Soc. 387, 1583 (2008).
\bibitem{jw1971} G. Jenkins and D. Watts, Spectral Analysis and Its Applications (Holden-Day, Merrifield, 1968; Mir, Moscow, 1971).
\bibitem{kaspi1994} V. M. Kaspi, J. H. Taylor, and M. F. Ryba, Astrophys. J. 428, 713 (1994).
\bibitem{lorimer2008} D. R. Lorimer, Living Rev. Relativ. 11, 8 (2008).
\end{thebibliography}
\end{document}